 \documentclass[final,authoryear,5p,times,twocolumn]{elsarticle}

\usepackage{amssymb,amsmath}
\usepackage{xspace,graphicx}

\journal{Astronomy and Computing}

\begin{document}

\begin{frontmatter}

\title{Hybrid polygon and hydrodynamic nebula modeling with multi-waveband radiation transfer in astrophysics}
\author[1]{W. Steffen}
\author[2]{N. Koning}
\address[1]{Instituto de Astronom\'{\i}a, Universidad Nacional Aut\'onoma de
M\'exico, Apdo. Postal 106, Ensenada 22800, Baja California, M\'exico}
\address[2]{Department of Physics and Astronomy, University of Calgary, Calgary, Canada}


%%\date{Received **insert**; Accepted **insert**}

%%\pagerange{\pageref{firstpage}--\pageref{lastpage}}

%%\maketitle
%%\label{firstpage}

\begin{abstract}
\noindent

We demonstrate the potential for research and outreach of mixed polygon and hydrodynamic modeling and multi-waveband rendering in the interactive 3-D astrophysical virtual laboratory \emph{Shape}. In 3-D special effects and animation software for the mass media, computer graphics techniques that mix polygon and numerical hydrodynamics have become common place. In astrophysics, however, interactive modeling with polygon structures has only become available with the software \emph{Shape}. Numerical hydrodynamic simulations and their visualization are usually separate, while in \emph{Shape} it is integrated with the polygon modeling approach that requires no programming by the user. With two generic examples, we demonstrate that research and outreach modeling can be achieved with techniques similar to those used in the media industry with the added capability for physical rendering at any wavelength band, yielding more realistic radiation modeling. Furthermore, we show how the hydrodynamics and the polygon mesh modeling can be mixed to achieve results that are superior to those obtained using either one of these modeling techniques alone.

\end{abstract}


\end{frontmatter}

\section{Introduction}
\label{intro}

For most objects that surround us in our terrestrial environment the radiation that makes them visible is emitted or reflected from its surface. In stark contrast, astrophysical nebulae are intrinsically volumetric phenomena and radiation may reach us from any part within them. The exception are stars, where optical radiation comes from a very thin layer that could be considered its surface. Therefore, the visualization of nebulae requires rendering techniques that differ profoundly from those used in software that model and animate scenes from our surroundings. In this introduction we briefly review different 3-D computer graphics approaches that are being used to visualize astrophysical nebula models for the purposes of scientific analysis and communication to the scientific community and the general public. We then present a technique that is new to astrophysics: a hybrid between polygon-based 3-D modelling and numerical hydrodynamics.

\begin{figure*}
  \begin{center}
  \includegraphics[width=0.48\textwidth]{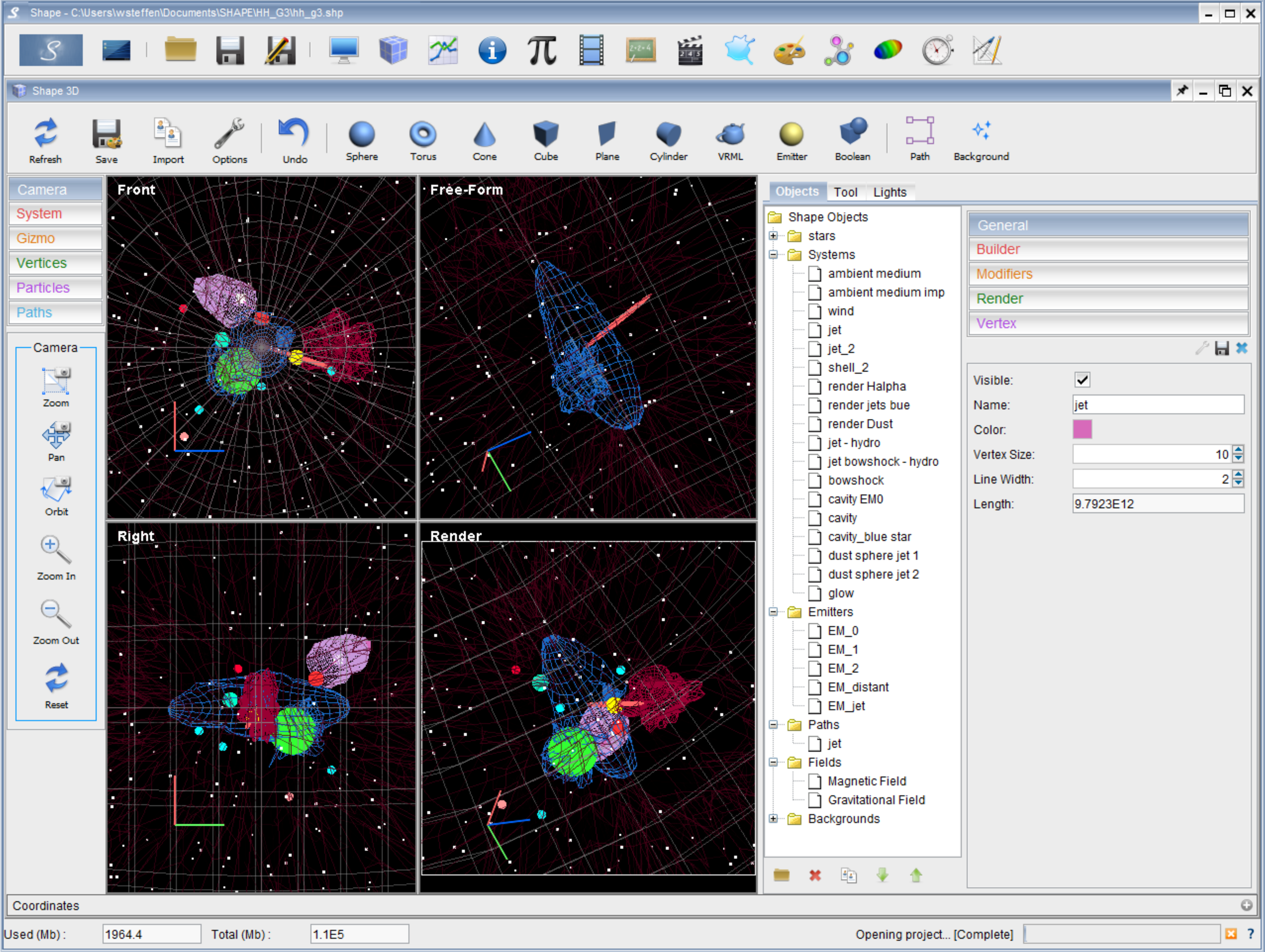}
  \includegraphics[width=0.48\textwidth]{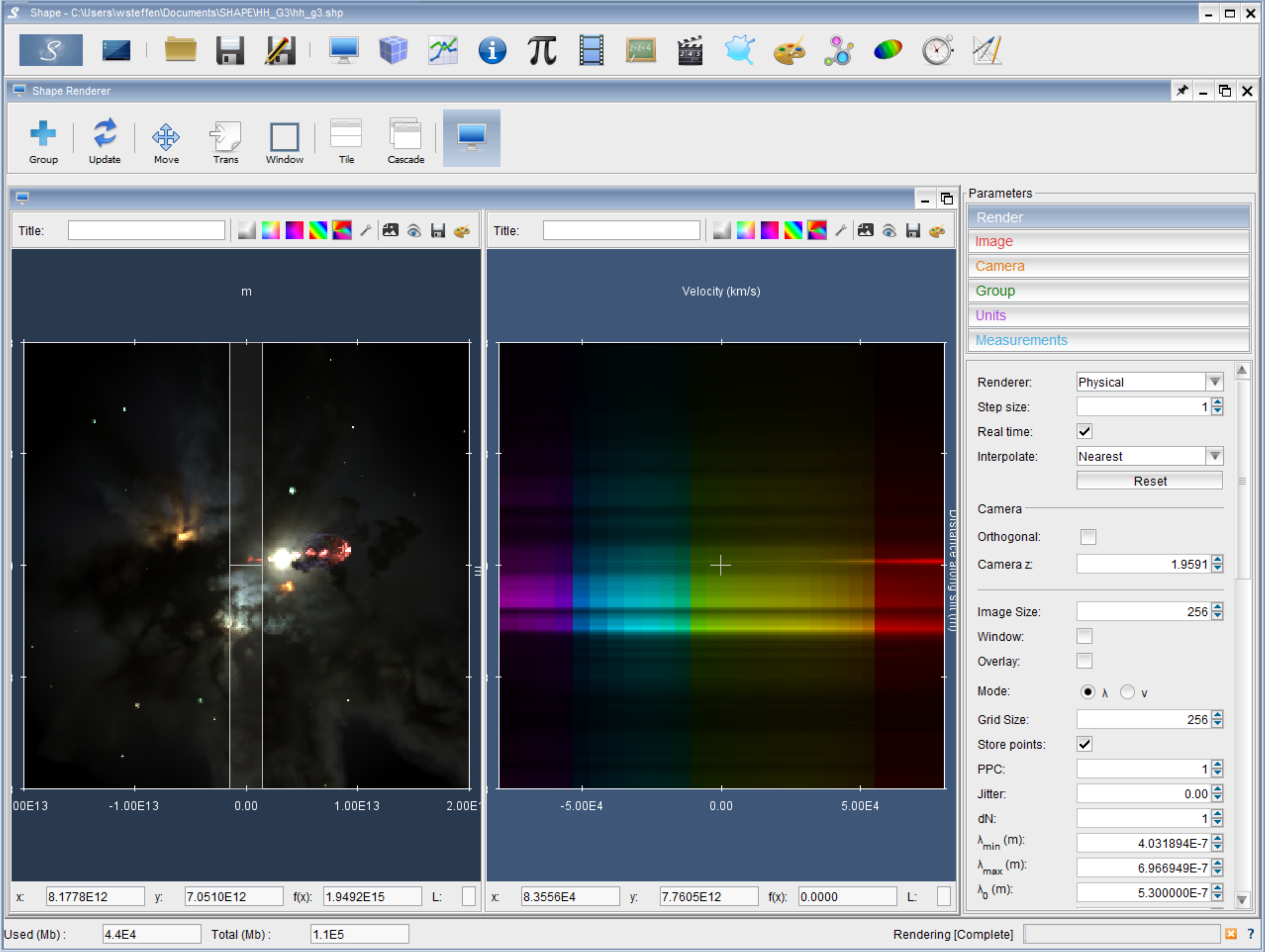}
  \end{center}
  \caption{Screenshots of the \emph{Shape} software show the 3D interactive modeling module on the left with a polygon mesh in the display. On the right is the same model after rendering, showing the image that includes dust scattered radiation from a central stellar continuum radiation source.}
  \label{shape_screenshot.fig}
\end{figure*}

The techniques of 3-D computer graphics is being widely used for the visualization and analysis of astrophysical 2-D and 3-D volumetric simulations (e.g. Kapferer \& Riser, 2008; Dolag et al., 2008; Kent, 2015; Pom\`erede et al, 2007). This kind of visualization is used to gain scientific insight from the simulations and for presentation purposes to the scientific community. Hassan \& Fluke (2011) review the usage of a variety of visualization techniques in different areas of astronomy over the previous two decades. They find that there is still a lot of room for development, in particular with the advent of big-data generation in observational and theoretical projects.

Grid and particle based (hydro-)dynamic simulations are visualized as volumetric renderings or using 3-D iso-contour surface extraction methods that then render the surface as a polygon mesh. Standard or custom transfer functions are used to obtain scientific insight from the visualization enhancing different local properties of the simulated fluid. A number of visualization packages offer programming or interactive interfaces to obtain the desired visualization and data analysis graphs. Usually the simulation and visualization software are separate entities and communicate via data that have been written to file.

A second type of application of computer graphics to astronomy is the generation of interactive 3-D photorealistic representations of astronomical objects on a general purpose graphics processing unit (GPGPU) (Magnor et al., 2005). The aim of these techniques is mainly educational. The advantage of this approach is that the user or viewer can interactively change the camera view in real-time. A disadvantage is that, due to the limitations of current GPGPUs, only a limited size and complexity of the scene can be visualized at the same time.

Similarly, non-realtime rendering of astronomical scenery for documentary productions is commonly done with standard commercial computer animation software. The advantages of this approach is that large scenes and very complex structure can be modeled and rendered. The disadvantage being that only a previously fixed camera path can be rendered and visualized, no user or viewer interactivity is possible. Furthermore, the rendering techniques are designed for terrestrial applications and the resultant radiation transfer does not correspond to what actually happens in space.

Occasionally, astrophysical simulation data are processed and modified to generate animations that can be used in documentaries or planetarium projections using purpose built rendering software on academic supercomputer systems (e.g. Nadeau et al., 2001). Nadeau (2008) gives a lively account of the potential benefits, challenges and problems that are likely to arise in such a project.

Generating special effects in the movie and television industry takes a different approach. Particle and grid based hydrodynamic simulations have been integrated into the existing interactive modeling frameworks that use polygon surface modeling and rendering. This allows not only the ability to combine the visual effects from the simulations with polygon based models of objects in the environment, but also their interaction (Yngve, O’Brien \& Hodgins, 2000).

%Initial and boundary conditions are set up in a similar interactive way as it is used to generate polygon objects. Polygon objects can be used as source or containers of the hydrodynamic simulations, which is only part of spatially larger scene (Figure \ref{shape_screenshot.fig}, left).

Since special effects applications of hydrodynamic simulations aim only at visually convincing behavior of the fluid motion after rendering, the accurate computation of physical quantities and their analysis a priori are not important and strong trade-offs regarding physical accuracy in favor of computational speed can be made that go beyond those permissible in astrophysics. Furthermore, computer graphics algorithms for gaseous fluid dynamics have concentrated on viscous and incompressible flows (e.g. Stam, 1999), while explosive compressible simulations have only recently been implemented to allow dynamical interactions of the gas with the solid objects in the scene (Yngve, O’Brien \& Hodgins, 2000).

In recent years Simon Portegies Zwart (Zwart et al., 2009) has been developing the Multipurpose Software Environment (MUSE) and its more recent variants AMUSE (Astrophysical Multipurpose Software Environment) and ABC-MUSE, in an effort to allow a systematic combination of a heterogeneous selection of simulation codes. It is based on a Python scripted environment, that the user customizes to use or include new codes; these include particle dynamics programs, radiation transfer or hydrodynamics codes, among others. While this scheme improves the way multiple simulation and visualization codes are applied together, it still requires extensive and detailed programming by the user as well as detailed knowledge of several codes for simulation, visualization and analysis.

In order to eliminate the need for user programming, Steffen et al. (2011) developed a fully integrated virtual astrophysical laboratory that began as an interactive morpho-kinematic modeling tool. To our knowledge {\em Shape} has been the first software to incorporate the interactive polygon mesh modeling technique specifically for astrophysical modeling (Steffen et al., 2011). In contrast to common astrophysical modeling, {\em Shape} tries to incorporate modeling, visualization and analysis tools in a single integrated program. {\em Shape} has been used for modeling the 3D structure of expanding nebulae (e.g. \citealt{GD12,LGSRR12}), novae \citep{R13}, and radiation transfer calculations in dusty pre-planetary nebulae \citep{KKS13} or quark-nova ejecta \citep{ouyed12}. \footnote{ The download and support site of the software is located at (http://www.astrosen.unam.mx/shape/)}

More recently, an Eulerian grid-based 3-D hydrodynamics module for compressible gas has been added to {\em Shape} and successfully applied to develop a new model for multipolar planetary nebulae (Steffen et al., 2013). Similar to what is customary in computer graphics (CG) animation software, the setup, visualization and analysis of the hydrodynamic simulations is fully integrated in the interactive 3-D modeling environment and requires no explicit programming by the user, which currently appears to make it unique in the area of astrophysics.

Another unique feature of this hydrodynamics module is that the simulation can be mixed with conventional polygon modeling,  thereby strongly enhancing the capabilities of each of them. In astrophysical research, this technique was used for the first time in Mehner et al. (2016). Hydrodynamic simulations, including externally computed ones, can be incorporated in each polygon model scene. This allows more realistic density and velocity fields, especially when shock surfaces are present, which is a key feature of astrophysical fluid dynamics. The presence of shocks precludes the usage of most software that is being used in terrestrial visual effects applications, which contain hydrodynamics modules that normally are based on incompressible Navier-Stokes equations (Stam, 1999).

In this paper we demonstrate the design and a few example applications for astrophysical science and visualization of nebulae that take advantage of the combined polygon and hydrodynamic modeling system with physical multi-waveband radiation transfer in {\em Shape}.

\begin{figure*}
\begin{center}
\includegraphics[width=0.96\textwidth]{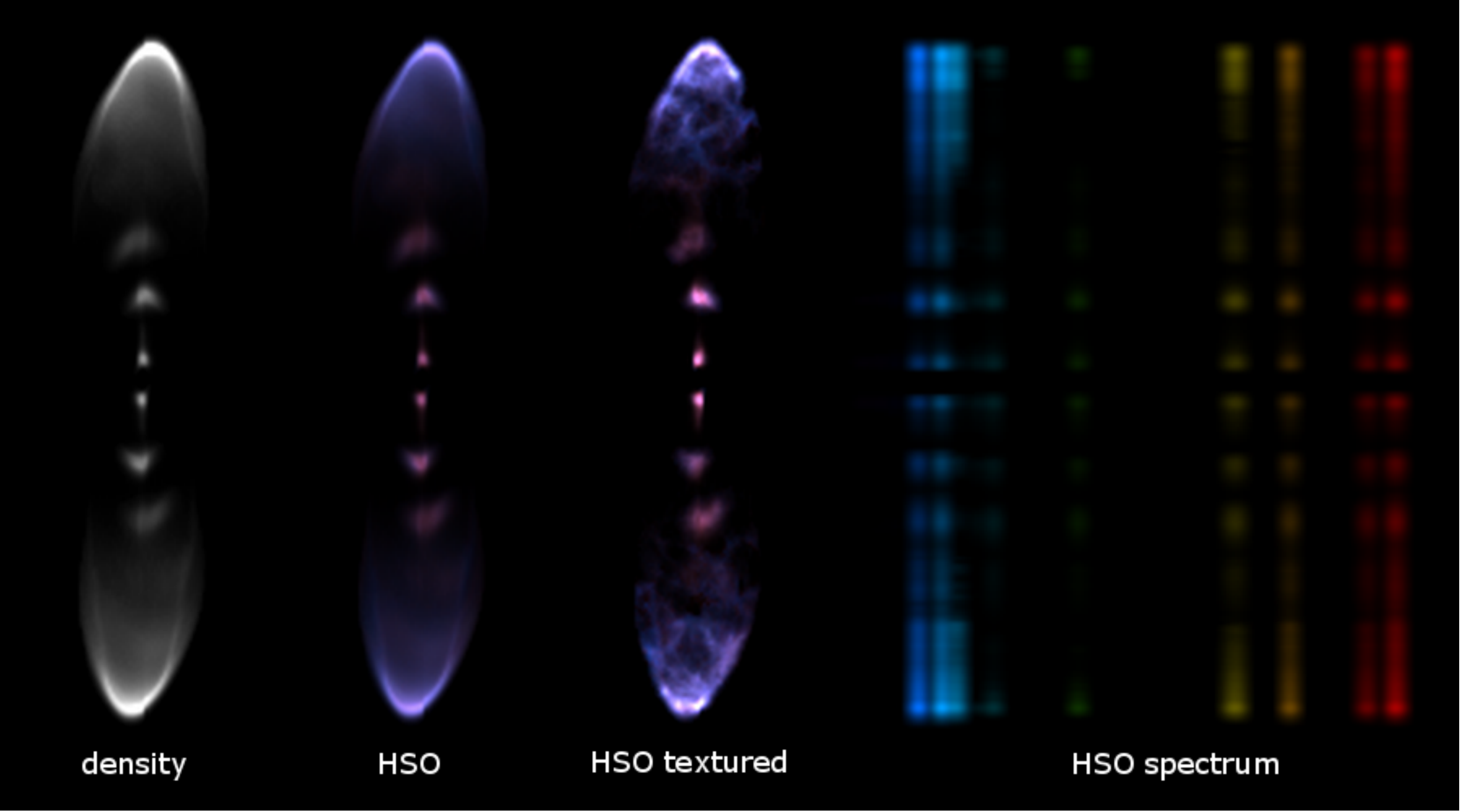}
\end{center}
\caption{On the left the rendering of a precessing and variable jet is shown as it appears in the grey-scale real-time preview during the hydrodynamic simulation in \emph{Shape}. The second image is the same simulation as rendered using atomic emission for { hydrogen, sulfur [++] and oxygen [+] (HSO)} with collisional excitation that depend on the local physical parameters. In the third image additional structure was applied through multiplying the density with a filamentary 3-D texture. Finally, the optical spectrum of the jet is show, with the vertical axis being the position along the jet axis. The intensity of the sulfur { and oxygen spectra have} been adjusted to be similar to that of hydrogen.}
\label{filteredhydro.fig}
\end{figure*}

\section{Programming language and parallelism}

\emph{Shape} was written in the \emph{Java} object oriented programming language. The language was chosen for its promise to be immediately compatible with any operating system, since astronomers use a wide variety of operating systems.

The 3-D interactive module was originally based on the \emph{Java3D} API, but has recently been changed to \emph{JMonkeyEngine} (Eden, 2014). The latter is more flexible than the former and has a large and active development community. Similarly, the user interface is in the process of being changed from the older \emph{Swing GUI} to the much improved and High-DPI aware \emph{JavaFX GUI}.

Many of the data processing and simulation calculations are able to use the multi-processor, multi-core and threading architecture of current CPUs based on shared RAM. This includes the radiation transfer, data transfer from the 3-D scene to the rendering grid, the rendering itself, the hydrodynamics and the interactive preview of the hydrodynamics. In \emph{Shape} we use the Fork/Join methods for parallel task distribution. An important property of this scheme is that threads that have finished, ask other threads that are still busy to move some of the pending tasks to the idle threads, thereby increasing the overall performance.

\section{Morpho-kinematic models from polygon meshes}
\label{polygonmodels.sec}
As is common in computer graphics animation software, in \emph{Shape} 3-D objects can be generated from surface meshes of triangular or quadrilateral polygons. These surface objects can then be rendered after assigning radiative properties either to the surfaces or the volumes that they enclose as a function of position and time. Since in astrophysical nebulae we are dealing with essentially volumetric processes, in \emph{Shape} the polygon meshes are used as volumetric enclosures that select regions to be rendered volumetrically (Figure \ref{shape_screenshot.fig}).

The construction of a complex structure begins with basic \emph{primitives}, such as spheres, tori and cylinders. These can then be changed by geometric \emph{modifiers} into very complex shapes. Virtually any imaginable structure can be built using a combination of these modifiers. The physical properties are then applied in the form of modifiers for density, temperature, velocity, etc. The radiative properties are set up in the form of what we call \emph{species}, which are kind of equivalent to the \emph{shaders} in computer graphics. The modifiers have tools that allow the user to vary the properties as a function of position by applying analytic functions, and manual key points in combination with 3-D procedural and image textures. This leads to very detailed and specific control over the spatial variation of physical and structural properties in a fashion that is analogous to non-astrophysical commercial animation software.

The voxelisation of the polygon meshes, i.e. their transfer to the 3-D regular grid used for rendering, is achieved by sending rays along the line of sight through each row of voxels. The ray is used to decide whether a grid voxel is inside or outside the polygon object. If inside, its physical properties are queried at this position and included as a property of this voxel. If several objects occupy the same voxel, the information is kept separate and is used individually in any radiation transfer operation. Multiple rays can be sent to improve accuracy by averaging the values for each ray. To speed up the process each object has a precomputed bounding box that is used to reduce the number of voxels to be checked whether they are within the volume of a particular polygon object. If a voxel is only partially inside the polygon volume, then the emission properties in the voxel are adjusted according to the fraction of volume that is inside.

In astrophysical applications the velocity field is particularly important as an observational quantity through spectroscopy. As a measurable diagnostic quantity and for the modeling of changing structure it is a key property. Similar to other physical properties, the velocity field can be applied as a modifier object and combinations of several can produce very complex fields. Visualization of the vectors in the interactive 3-D view ports help to correctly set up the velocity field or visualize it for publication.

By default physical units are in the international MKS-system, i.e. lengths are measured in meters, mass in kg and time in seconds. Physical and angular distances in space and on the sky can also be handled in units that are more common in astronomy, such as parsec or AU and arcsec, arcmin, respectively, to mention just a few.

\section{Hydrodynamic models}
\label{hydromodule.sec}

In contrast to the numerical hydrodynamics schemes that are in general use in conventional 3-D animation software, the code in \emph{Shape} is able to simulate the supersonic flow that is common in astrophysics with the formation of shocks. This code has been described in \cite{SKE13}, where it was used to simulate multipolar planetary nebulae. It uses a uniform cartesian grid and a van Leer flux-vector splitting scheme.

The setup of a simulation is done using 3-D polygon objects that determine the initial and internal boundary conditions. Instead of being rendered, these objects get assigned a \emph{hydro} modifier that identifies them as objects to be used in a hydro simulation. Other objects are ignored for the setup of a simulation. The \emph{hydro} modifier has three settings that identify the object type. First, it can be an initial condition for the background that applies to the full simulation domain (here the polygon mesh is a dummy structure). The second option sets the object to be added only as an initial condition. The third option sets the objects to an internal boundary condition, that is reapplied at every time step, such as stellar winds or jets that continuously inject new material. The properties of the internal boundary objects can be variable in time. These variations are set up through the \emph{animation module} and may include changes in structure or the physical parameters.

Rendering of the hydrodynamics is done via dummy polygon objects in the 3-D module. To access the hydro data there are two options, either directly from the current simulation in the hydro module or loading the hydro grid from a file on disk. The latter allows more than one hydro simulation to be included in a rendering, making it possible to simulate local processes separately at higher resolution than the overall spatial domain allows at render time.

To reduce memory requirements rendering the simulated data can be filtered using spatial and physical criteria such as velocity, density, temperature or pressure to extract features of interest. From a single simulation, several different features can be extracted using a dummy polygon object for each of them. Each of these objects can be rendered with its own single or multiple {\em species}, which determine the radiative properties such as emission, absorption or scattering coefficients. These coefficients are then calculated from density, velocity, temperature and dust properties, if applicable.

In Figure (\ref{filteredhydro.fig}) we show the results from a hydrodynamic model of a precessing bipolar jet with variable density and velocity with the integrated density shown in grey-scale on the left. Using the atomic species from the Physics Module of \emph{Shape} that depend on the local physical properties of the gas, a visually interesting color image is produced (second image). Additional realism is achieved through multiplying the density structure with a 3-D noise texture prior to applying the atomic species (third image). Additional scientific and educational insight can be obtained displaying the resulting optical spectrum, showing the different spectral lines as their relative intensity changes along the jet features.

The spatial resolution of the hydrodynamic simulations presented in this work is $250^3$ voxels. The adiabatic index ($\gamma = 1.1$) was chosen such that an approximately isothermal flow was achieved without explicitly computing the cooling as a function of temperature (\citep{SKE13}).

\section{Mixing volumetric polygon and hydrodynamic models}
\label{mixingpolygonhydro.sec}

The main novelty that we present in this paper compared to other astrophysical simulation and visualization software is the possibility to mix polygon objects with one or more hydrodynamics simulations in the same interactive computing framework. To the best of our knowledge, in astrophysics \emph{Shape} is currently the only system to allow this in a straightforward way.

The rendering framework of \emph{Shape} is based on the generation of a cubic voxel grid with its data provided by any renderable polygon type of object and light source present in the 3-D module.

In \emph{Shape} the hydrodynamics does not dynamically interact with the polygon objects that have not been incorporated as initial conditions of the hydrodynamic simulation or later during the simulation as part of the environment. To avoid significant physical inconsistencies regarding the presence of polygon objects during rendering, the user must take care not to use polygon objects that would have been substantially modified by the hydrodynamic flow and vice versa.

\section{Radiative transfer and \emph{shading}}

A fundamental problem with most 3-D animation packages that were not designed for astronomical modeling and visualization is that the output images from volumetric models are produced in a way that differs not only quantitatively, but also qualitatively from physical radiation transfer. Their \emph{shaders} are designed to produce a realistic view of solid, possibly refractive or semitransparent terrestrial materials. A key feature is that the transparency or opacity of translucent materials are linked to the emission. The result is that completely optically thin objects, such as many nebulae, become invisible. Directional scattering from dust particles are not handled properly. This makes the ubiquitous reflection nebulae a challenge to accurately visualize.

Furthermore, the spectral description of the radiative properties is limited to the ARGB system, with values for A (transparency), R (red), G (green) and B (blue) color channels. A more detailed description of the spectral properties as a function of wavelength is not available. This precludes the modeling of line spectra. In particular the Doppler-effect can not to be taken into account, which is fundamental for the kinematic analysis of astrophysical nebulae.

\subsection{Render pipeline}
\label{render.sec}

The render pipeline is based on the data in a cubic voxel render grid that contains all the necessary radiation information. The information consists of emissivity, opacity and scattering properties for each spectral channel. Each voxel location is probed in the individual objects of the scene maintaining their separate identity within the grid. When keeping all the information is not necessary, this mode can be switched off to save memory and computing resources. Voxels that receive no information from the 3-D objects are discarded to save memory. Empty voxels are not used for rendering.

\begin{figure*}
\begin{center}
\includegraphics[width=0.48\textwidth]{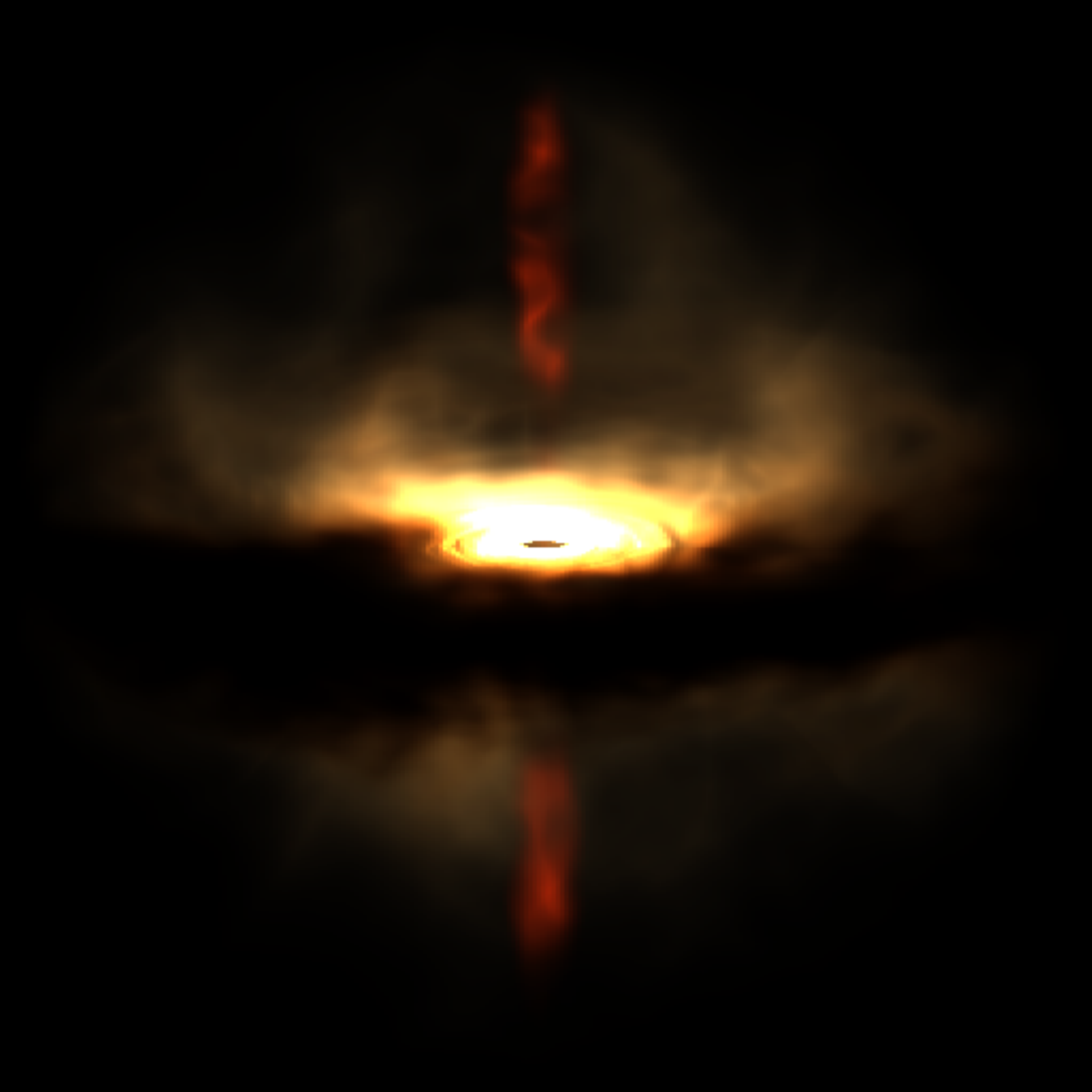}
\includegraphics[width=0.48\textwidth]{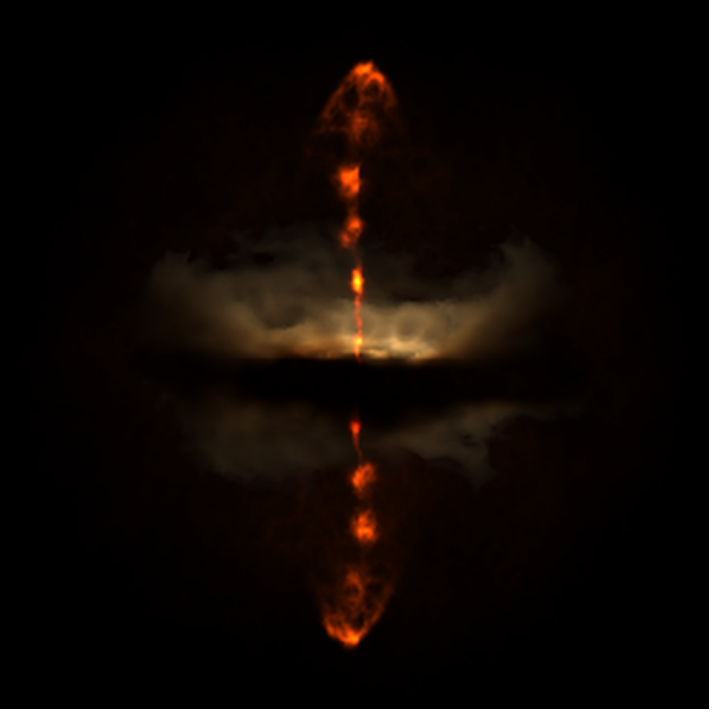}
\end{center}
\caption{Models of a proto-planetary disk are shown with radiation from the central star scattering of the surrounding dusty disk. The version on the left has a polygon model of an emission line jet, whereas on the right side the hydrodynamically computed emission line jet from Figure (\ref{filteredhydro.fig}) was incorporated in the polygon model of the dust disk.}
\label{polygonhydromix_ppd.fig}
\end{figure*}
\noindent
Render pass 1: The data from each 3-D object are passed to the render grid. \\
Render pass 2: Rays are cast from the back of the grid to every pixel on the camera plane, where the final intensity is recorded for each spectral channel. Multiple rays per pixel at random positions within the perimeter of the pixel can be cast to improve the accuracy of the result. If required, during the ray casting, radiation transfer for scattering processes is done from each light source to the currently processed voxel.
Each image pixel maintains the full spectral information which can be passed on to other modules for further processing in pass 3. \\
Render pass 3: For spectral output, such as position-velocity diagrams, channel maps or line profiles, the emission is sent to the corresponding bins in the spectral images and line profiles.

The rendering of images and spectroscopic output in the astrophysical software \emph{Shape} is produced using a very general numerical scheme for the radiation transfer. The number of spectral channels to be computed can be set by the user in order to compare the model with observations. A raw image is rendered using a ray from the back towards the camera taking into account the opacity for every spectral channel at each voxel position. In this raw image the complete spectral channel information is conserved, i.e. it actually is a 3-D data cube with the \emph{xy} plane being position information and the \emph{z}-direction being the spectral domain. The final color image is then formed by adding the individual channels according to three wavelength ranges each of which represents an RGB color.

%DELETED  XXX
%Opacity is taken into account for each spectral channel along the ray that integrated the radiation from the back towards the image plane.

Four different rendering schemes are used to optimize the rendering speed depending on whether the model is optically thin or contains light scattering from point sources. Details of the theory of radiation transfer and the numerical algorithms used can be found in a special document on the software´s website \footnote{ (http://www.astrosen.unam.mx \\ /shape/downloads/files/Physics\_in\_SHAPE.pdf)}. { The detailed radiative properties of a gaseous or dust materials are set up in the \emph{Physics Module} discussed in subsection \ref{physicsmodule.sec}. }

A \emph{basic} renderer adds all the emission along a single ray for each pixel, ignoring opacity. This renderer is meant for fast rendering of optically thin objects. It produces images, P-V diagrams, channel maps and 1-D spectral line shapes.

The \emph{physical} renderer applies full radiation transfer according to the needs of the different components in the model. This may include dust scattering or molecular CO line transfer in the large velocity gradient approximation. Dust scattering calculations include a two step transfer from every voxel to each of the light sources that are present in the scene and then the transfer from the back to each pixel on the image plane. For this rendering a 3-D cubic cartesian voxel mesh is set up with all necessary radiative and physical properties that have been transferred from the 3-D scene prior to the calculation of the radiation transfer and rendering. Once the regular 3-D voxel mesh has been computed, the rendering of different view points, e.g. to generate a movie with the camera flying around or into a nebula, can be highly accelerated by only casting rays from the back to the front onto each pixel.

In scenes where no scattering of light needs to be calculated the 3-D regular grid is not necessary and a direct ray-casting from the back to the front can be applied right away that takes into account emission and opacity. This allows very high resolution images and spatially resolved spectra to be computed, since only the line of sight contributes to the emission that reaches a pixel. Using parallel computing accelerates the calculation and many pixels can be computed simultaneously with this high definition (HD) rendering mode.

\subsection{ Setup of radiative properties}
\label{physicsmodule.sec}

{

The radiative properties of a material as a function of wavelength, density, temperature, pressure, etc., is set up in the \emph{Physics Module} in the form of what we call \emph{species}. There are a variety of species that calculate emissivity, opacity and scattering in different ways. The \emph{Default Species} has emissivity proportional to the density and no absorption or scattering. This species is assigned to all objects at the time of creation for quick experimental modeling.

For more advance modeling, species other than the \emph{Default Species} are available.  For example, a \emph{Custom Species} allows the user to set emissivity, opacity and scattering properties as a function of the local physical state of the gas or dust in various ways. They can be given using an analytical function of these properties or as a data driven function of wavelength or a combination of both. The local material properties are set up explicitly as a function of position and, if necessary, of other local properties. Note that temperatures are currently \emph{not} computed self-consistently from the radiation transfer, but set explicitly as a function position of other properties. It is also possible to include results from other external codes in the custom species; for example cross sections computed by DustEM (Compi\`{e}gne et al., 2011).

Another example of a more advanced species is the \emph{Dust} species which calculates the emission, absorption and scattering coefficients using either the full Mie theory or an approximation to it (see section 1.5 from the aforementioned document \emph{Physics in SHAPE}.  The user can select a preset dust species (amorphous carbon, silicates or graphites) or they can specify the grain sizes manually using the MRN (Mathis, Rumpl \& Nordsieck 1977) power law.

The \emph{Atomic} species uses either the Chianti (Dere et al., 2009) or Kurucz (Kurucz \& Bell, 1995) data bases (the user can choose either) to calculate the bound-bound (see for example Eq 5.10 and 5.11 from Kwok, 2007), and bound-free (Kwok, 2007) coefficients from an (ionized) atom. The free-free (or Bremsstrahlung) coefficients can also be calculated with the \emph{Atomic} species, as well as collisional excitation, LTE and non-LTE (using two level statistical equilibrium approximation) emission/absorption, and Thomson scattering. Recombination from photo-ionization, however, is not currently included.

Additional radiation effects can be computed in a simplified way including photo-ionization, cyclotron and synchrotron radiation. However, these are not relevant for the results and core message of this paper and will not be discussed here.

} % end bf physics module

\subsection{ Ray tracing}
\label{raytracing.sec}

{
\emph{Shape} uses a ray tracing algorithm to perform radiative transfer. For each pixel in the output image, a ray is created which holds the intensity for the pixel in question and ``extends'' from the back of the scene to the front. The algorithm steps through the scene along the direction of each ray in set increments, ds, and adds/subtracts emission from the ray along the way. At each step, the emission (j), absorption ({$\rm k_a$}) and scattering ({$\rm k_s$}) coefficients at that location are used in conjunction with the equation of radiative transfer to determine the intensity of the ray after the step. Once the ray has passed through the scene, the color and intensity of the pixel is determined and set in the final image.
} % end bf ray tracing

\subsection{ Scattering}
\label{scattering.sec}

{
If point-sources of light are present in the scene, as they are in this work, the above algorithm is slightly modified.  At each step along the ray a second ray is cast from the light-source towards the current position in the scene. The ray from the light-source has an initial spectrum, defaulted to a Plank-spectrum to simulate star-light. This second ray undergoes the same radiative transfer as described above. The intensity of this second ray is then used in conjunction with the scattering coefficient to determine the resulting intensity of the first ray. A detailed description of the algorithm used in Shape can be found in the aforementioned document \emph{Physics in SHAPE}.
} % end bf Scattering

\subsection{Post-render real-time interactive GPU visualization}
\label{gpuvisualization.sec}

An emerging application for volumetric data of astrophysical models are real-time interactive visualizations in planetariums and scientific presentations. \emph{Shape} has output capabilities that allow the volumetric models to be visualized with external software, generally providing ARGB channels. If desired, a higher number or different channels, including invisible bands such as infrared, x-ray or radio can be output.

These visualizations can be rendered interactively on graphics cards (GPU). The volume grid in \emph{Shape} can be exported as image slices, which are then loaded as textures in the real-time visualization system at resolutions that depend strongly on the video RAM capabilities of the system. Other output modes can be set up for other external volumetric visualization software.

\begin{figure*}
\begin{center}
\includegraphics[width=0.32\textwidth]{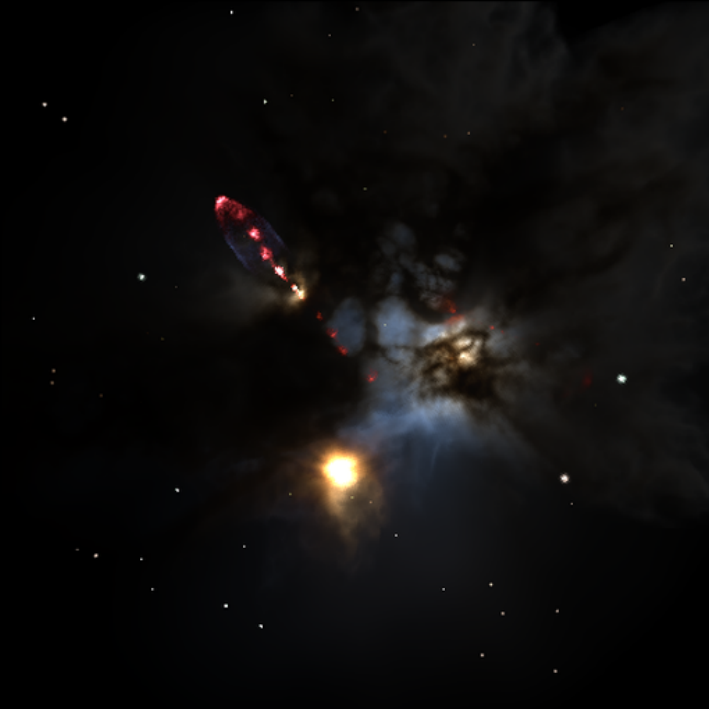}
\includegraphics[width=0.32\textwidth]{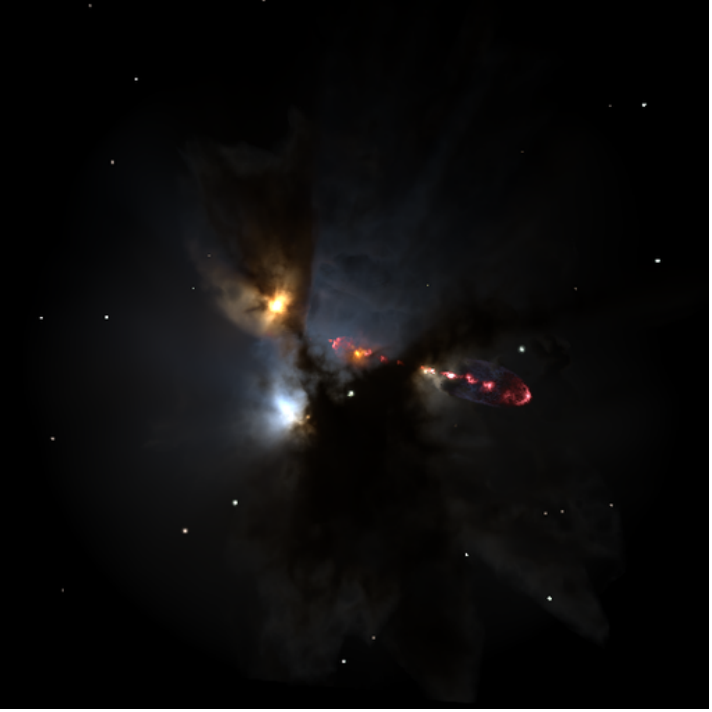}
\includegraphics[width=0.32\textwidth]{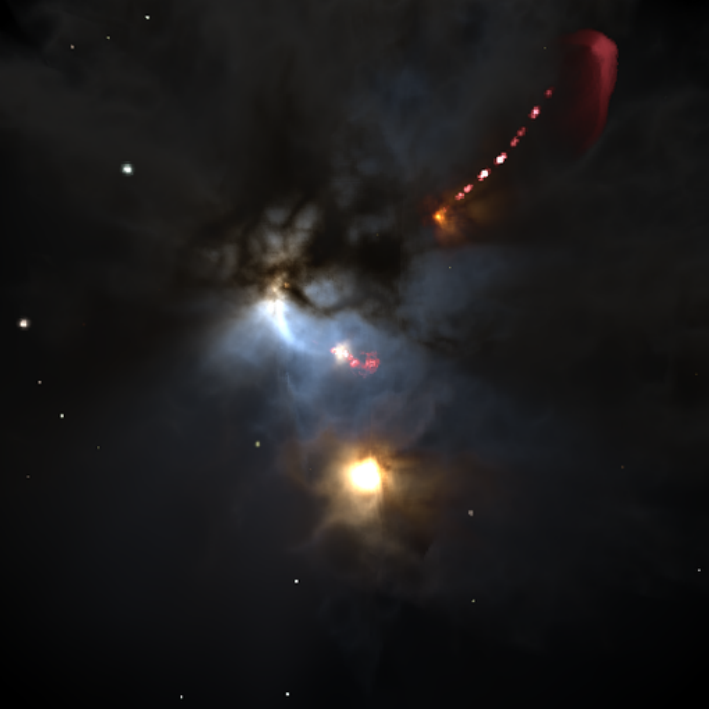}
\end{center}
\caption{The rendering of a dusty star forming region with four different embedded stars. Optical images from three different viewing angles are shown. The stellar radiation is scattered off the surrounding dust. The region includes two jets, one modeled with a polygon mesh and a second that was computed hydrodynamically. The emission of both jets is a single emission line at the wavelength of H$_\alpha$. The images where rendered at $450^2$ pixels. On the right the prominent jet is from the hydrodynamic simulation, while for the jet in the image on the right a polygon object was used.}
\label{polygonhydromix_nebula.fig}
\end{figure*}

\section{Results}
\label{results.sec}

While a polygon based 3-D model allows highly complex and specific modeling of particular astrophysical objects, a realistic velocity field is often not easy to determine from observations or set up in \emph{Shape}. In such cases a hydrodynamic simulation can be a quick solution for a more realistic model, both for the structure and the velocity field. In Mehner et al. (2016) we demonstrated for the first time that a mixed model using both polygon and hydrodynamic components can provide a more convincing overall scientific model than a pure polygon model.

In this section we describe the results of combining polygon model objects with a hydrodynamic simulation component for scientific analysis and photo-realistic visualization. Furthermore, we demonstrate the visualization of nebulae in different spectral wavebands.

\subsection{Dusty polygon and 3-D texture objects with a hydrodynamic jet}
\label{polygonhydromix.sec}

We show that such a mixed model can improve models intended as photo-realistic representations of astrophysical nebulae that are meant for outreach work. Figure (\ref{polygonhydromix_ppd.fig}) shows on the left an earlier version of a proto-stellar dust disk with a jet emerging from the stellar neighborhood, where both, the dusty disk and the emission line stellar jet were modeled using polygon objects with a 3-D noise texture. The dust disk is illuminated from a star at the center and scatters the continuum stellar light. The light from the jet is a narrow emission line (H$\alpha$). On the right, the polygon jet has been replaced with a hydrodynamic simulation of a precessing episodic jet that was generated with the hydrodynamics module in \emph{Shape}. The filamentary details of the jet structure were achieved by multiplying the simulated density field with a 3-D filamentary noise structure, yielding a more natural appearance of the disk-jet system than the original full polygon model. In section \ref{longslit.sec} we show how such a simulation can also be set up for spectroscopic analysis as a function of orientation of the object towards the observer.

In Figure (\ref{polygonhydromix_nebula.fig}) the same hydrodynamic jet has been incorporated in a physically large star forming region where four stars with different luminosities and temperatures illuminate a dust dominated nebula that reflects their differently colored light. The region also incorporates a second jet with a bowshock that was created using polygon structures. For the hydrodynamic jet, a non-linear emission species was set up that uses the density and temperature structure to differentially color the jet and dense regions of the bowshock in red, while the cooler low-density wings of the jet appear blue, signaling the differences in the physical states of the gas in the different regions of the flow.

No velocity field was assigned to the polygon structures. However, since the hydrodynamic jet automatically incorporates the velocity field, the model can be used to demonstrate the effects of complex opaque dust structures on the spectroscopic appearance of high velocity flows that emerge from the dense regions of dusty star formation. This is shown in section \ref{longslit.sec}.

\subsection{Proto-planetary disk from the optical to the infrared}
\label{waveband.sec}

As we have shown above, the application of hydrodynamic simulations substantially improves the structural appearance and the kinematic properties of astrophysical flows that are part of a scene that is otherwise composed of polygon objects. The second important subject for achieving realistic visualizations of astrophysical objects that we address in this paper are physical radiation source and transfer computations.

Astrophysical insight more and more relies on multi-wavelength observations that may go from the high-energy Gamma and X-ray photons to the long wavelength radio spectral range. In scientific publications, and press releases for the general public, often there are images taken in optically invisible wavelength bands. Sometimes these are superimposed on optical data in a single image. For the general public such images are not easy to interpret. It may not be clear why the objects change so much at different wave bands. Visualizations of the appearance in a continuous transition from the more familiar optical appearance to the infrared or beyond is likely to help in the understanding of long-wavelength images.

A key confusing factor is that the assignment of color to the optically invisible emission is basically arbitrary, often not even conserving the order with respect to the colors in the optical image, such as X-rays that are not shown as blue or infrared that is not red. How the colors have been mapped necessarily requires an explanation.

In addition to a verbal explanation of the reasons for the difference in appearance, it might be helpful to show a continuous sequence or movie of how an object changes as the wavelength band is shifted continuously through the electromagnetic spectrum. We demonstrate this effect by modeling the emission of a protoplanetary dust disk with an emission line jet.

\begin{figure*}
\begin{center}
\includegraphics[width=0.95\textwidth]{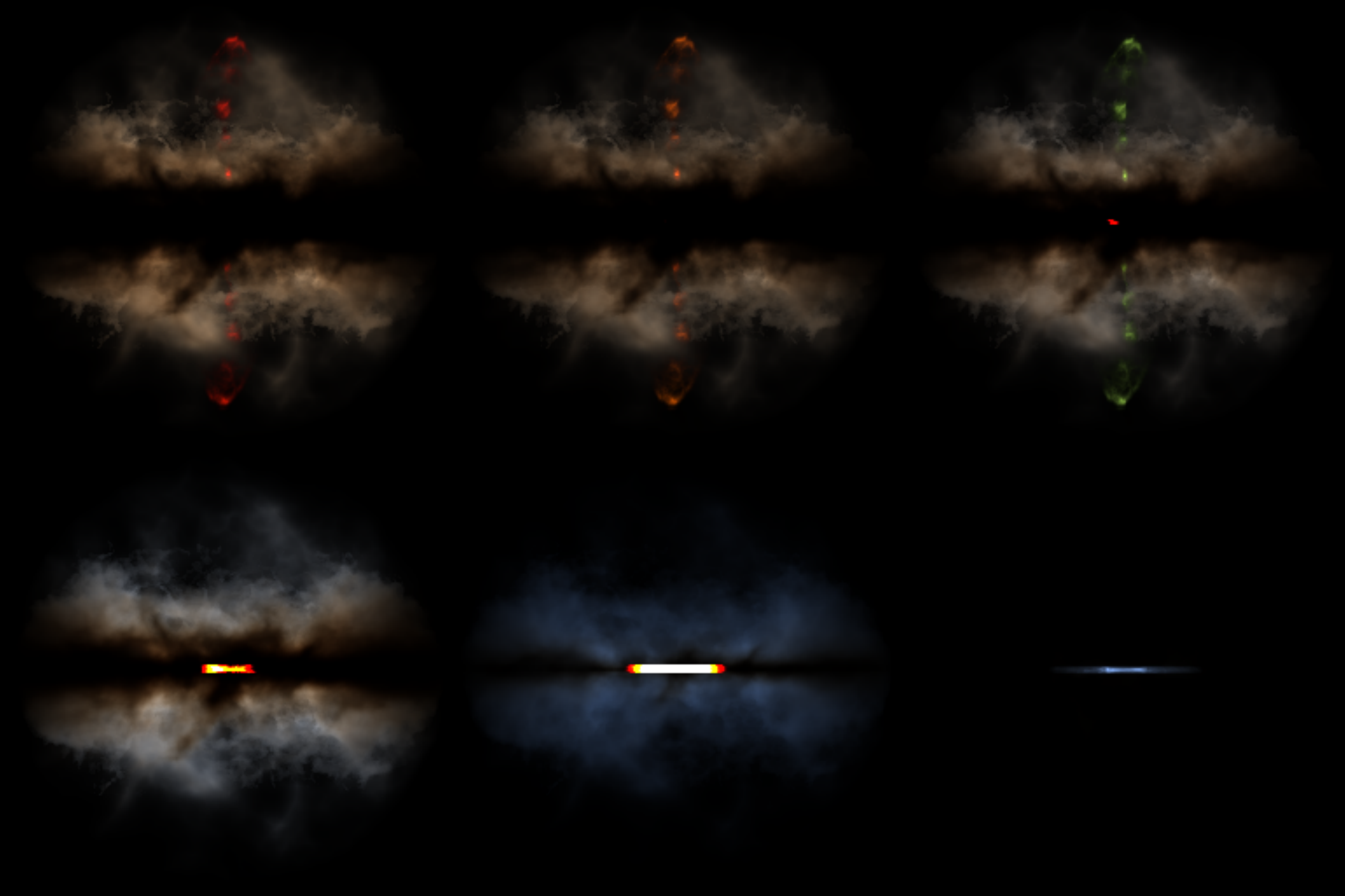}
\end{center}
\caption{The wavelength range in the rendering setup for model from Figure (\ref{polygonhydromix_ppd.fig}) was modified to show the appearance of the nebula going from the optical ({ top left}) to the far infrared ({ bottom right}). The ratio between the maximum and minimum wavelength remains constant. The color mapping from blue to red does also remain the same over the full range of rendered wavelength.}
\label{waveband_ppd.fig}
\end{figure*}

In Figure (\ref{waveband_ppd.fig}) and the accompanying movie, the change in appearance of the proto-planetary disk is shown as the wavelength changes from the optical range (400-700~{\rm nm}) to the far infrared (0.1-0.2~{\rm mm}). The same continuous rainbow color map is assigned to all rendered wavelength ranges, blue at the smallest and red at the largest wavelengths. As the waveband changes, the visible structure changes, too. In the infrared the opaque dust lane becomes optically thin and reveals the thin inner disk. As the wavelength of the synthetic image increases, the scattered light weakens strongly and the optically thin thermal emission from the dust begins to dominate.

The color change and disappearance of the jet in the IR is particularly striking. Its emission has been represented by a single H$\alpha$ spectral line and therefore represents a very narrow and pure color. As the observed waveband continuously moves from the full optical range into the infrared, the single spectral line of the jet adopts false colors that include all the colors of the rainbow. As soon as the observed waveband moves beyond H$\alpha$, the jet disappears and can not be detected. A real jet would of course still emit some spectral lines in those wavebands. If desired, this can of course be incorporated in the emission model for the jet. While the mixed color of the continuum emission from the dust also changes from a whiteish hue to a blueish one, it never shows the pure colors of the rainbow.

At the center of the nebula is a stellar object with a Planck spectrum of temperature 4000 K and $8.8\times 10^{31}$~\rm{Watt} luminosity. The dust has a power-law particle distribution with an index of 3.5 { (Mathis et al., 1977)} and minimum and maximum particle size of $5\times 10^{-9}$~{\rm m} and $2.5\times 10^{-7}$~{\rm m}, respectively, with single isotropic scattering for all particle sizes. The physical size of the image is 266~{\rm AU}. The temperature of the dust is constant 300~{\rm K} for the thin inner disk and 50~{\rm K} for the rest of the nebula. The extinction properties are calculated as an approximation to Mie theory.

The jet was simulated hydrodynamically and scaled to fit into the computational region for the protoplanetary disk. The emission of the jet was set to be a single spectral line at the wavelength of H$\alpha$ scaled in brightness to be comparable with the brightness of the scattered light from the dusty disk.

In contrast to seeing images from separated wavebands, the image sequence and the corresponding movie highlight and help in the understanding of several noteworthy effects. These are:
\begin{itemize}
  \item The optically thick dust becomes transparent in the IR bands
  \item As the wavelength range changes, the interpretation of the color changes (from reflection to thermal emission)
  \item The color of the spectral line changes and disappears in the IR
  \item Scattering becomes negligible and thermal dust emission takes over
  \item The changing color behaviour of continuum radiation is very different from that of spectral lines
\end{itemize}

A key insight that the visualization with continuously changing spectral range shows is the quite sudden transition from optical thickness to full transparency in the infrared, when the wavelength becomes significantly larger than the grain sizes. This is of course particularly important for totally embedded objects. It clarifies one of the reasons why modern ground-based and space astronomy is more and more dominated by infrared observations that are able to peer deep into dust enshrouded regions or far into the universe where optical emission has been shifted to the infrared by the cosmic redshift.

\begin{figure*}
\begin{center}
\includegraphics[width=0.48\textwidth]{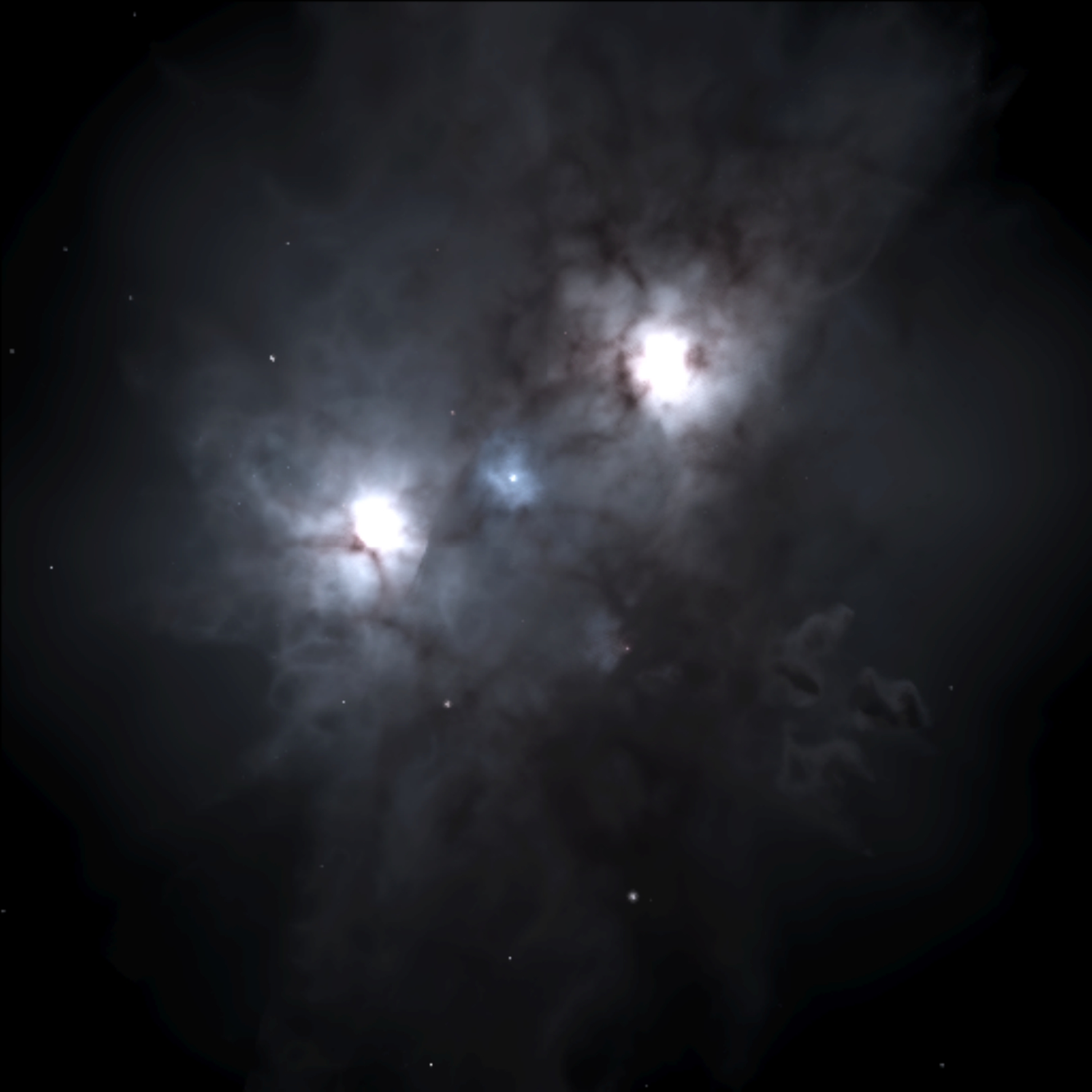}
\includegraphics[width=0.48\textwidth]{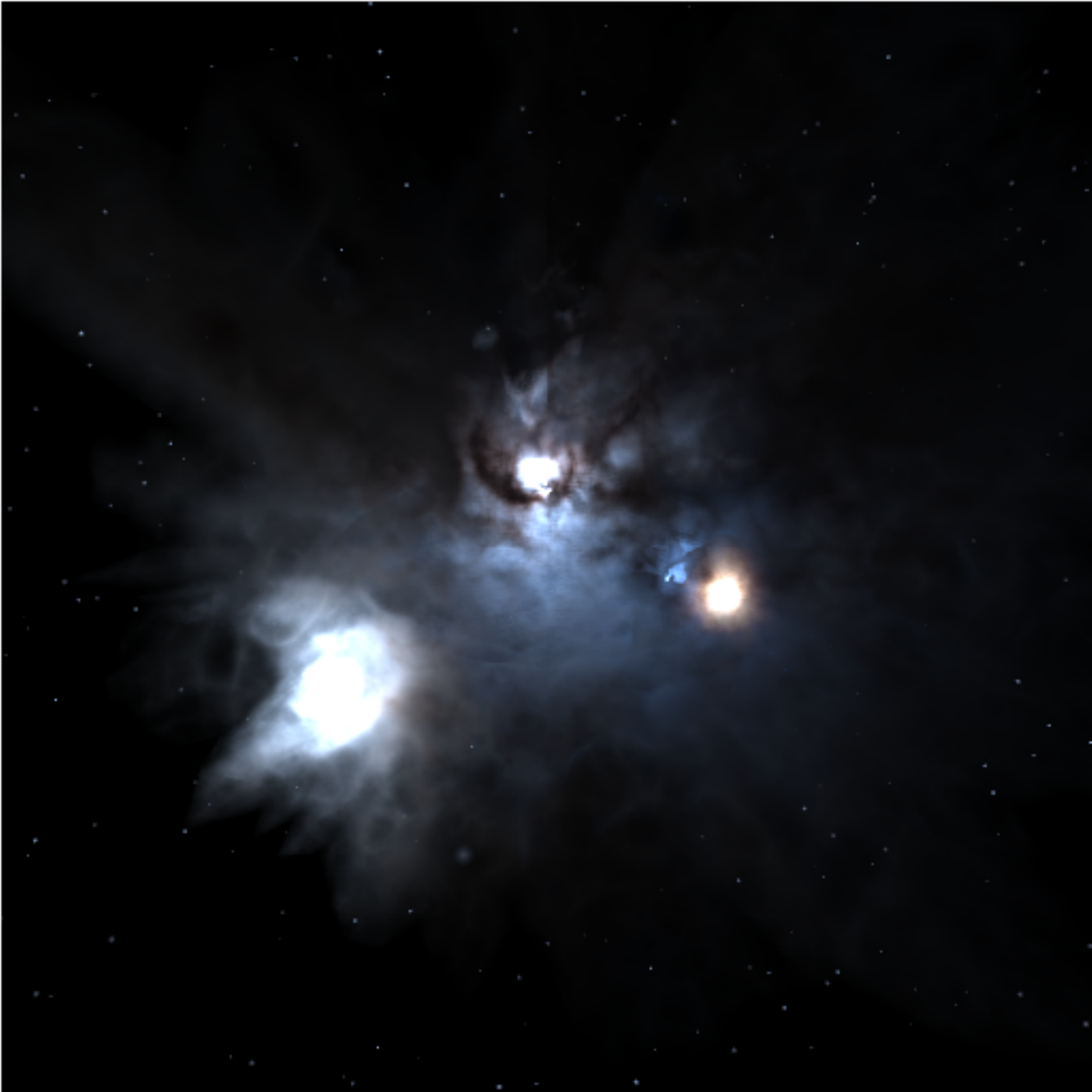}
\end{center}
\caption{The wavelength range in the rendering setup for model from Figure (\ref{polygonhydromix_nebula.fig}) was modified to show the appearance of the nebula in the infrared J,K and L infrared bands (blue, green, red color channels respectively). Two different viewing directions are shown. It demonstrates how the opacity of the cloud is reduced letting the embedded and background star shine through. The emission is still dominated by reflected star light. The jets disappear, because infrared emission for them is very low and has not been included in the model.}
\label{waveband_nebula.fig}
\end{figure*}

The transition from continuum scattered light to continuum thermal radiation of the dusty disk is not as obvious as the other properties discussed above. Only the change from an optically thick obscuring dust band to an optically thin radiation hints to a dramatic change, but is of course not enough to clearly notice the change in the qualitative character of the radiation.

To facilitate the understanding of the continuous changes that occur, the animation and image sequence can be complimented with a color bar that shows how the spectral range shifts from the optical to the infrared and sub-millimeter region.

\subsection{Star forming region}
In this section, we show a synthetic infrared observation of the dusty star-forming cloud for different wavebands (Figure \ref{waveband_nebula.fig}). While the optical image from Figure (\ref{polygonhydromix_nebula.fig}) shows the dominant effect of obscuration, in the infrared the dust is more transparent and the scattered stellar radiation resembles that of an emission nebula, except that it is less colorful since the radiation is a broad continuum rather than narrow spectral lines. Currently \emph{Shape} can not compute a self-consistent temperature distribution for the dust, based on the radiative heating by the stars in the region. However, if one can estimate an approximate temperature distribution as a function of position, it can be applied to the dust. This would allow one to extend the wavelength range further in the infrared, where the thermal emission of the dust becomes dominant. This effect was demonstrated for the somewhat simpler geometry of the proto-planetary disk in Section \ref{waveband.sec}.

The key scientifically relevant effect of the dust on the jet and bow-shock is that it \emph{locally} changes the spectral line profile, line-ratios and position-velocity diagram due to the spatially strongly varying absorption. Spectral line ratios change because of the decreasing absorption from small to large wavelength (the effect of \emph{reddening}). In the following section we provide examples of how the models can be used to demonstrate the influence of the dust absorption on the shape of position-velocity diagrams and spectral line profiles.

\subsection{Longslit spectroscopy}
\label{longslit.sec}

Some kinematic properties of astrophysical gas flows can be measured by way of the Doppler-effect in high-resolution spatially resolved spectroscopy. In the optical the most common method is long-slit spectroscopy, although integral field units that produce three-dimensional data cubes and channel maps are becoming more and more common. \emph{Shape} is also routinely used to model channel maps for radio observations. In this section we demonstrate that a model that mixes mesh with hydrodynamic simulation and has primarily been generated for visualization, can straightforwardly be adapted for spectroscopic modeling.

The possibility to mix hydrodynamic modeling with polygon mesh models allows a combination of the realistic velocity fields that result from a hydrodynamic simulation to be complemented with additional structure that is being observed but might not be part of the simulation or might be outside the simulation domain. This allows one to adjust some properties of a hydrodynamic model to a specific object, which otherwise is more difficult to obtain. Naturally, this may not be self-consistent modeling, such that the author has to apply special care not to introduce unreasonable properties of the added polygon structures.

\begin{figure*}
\begin{center}
\includegraphics[width=0.96\textwidth]{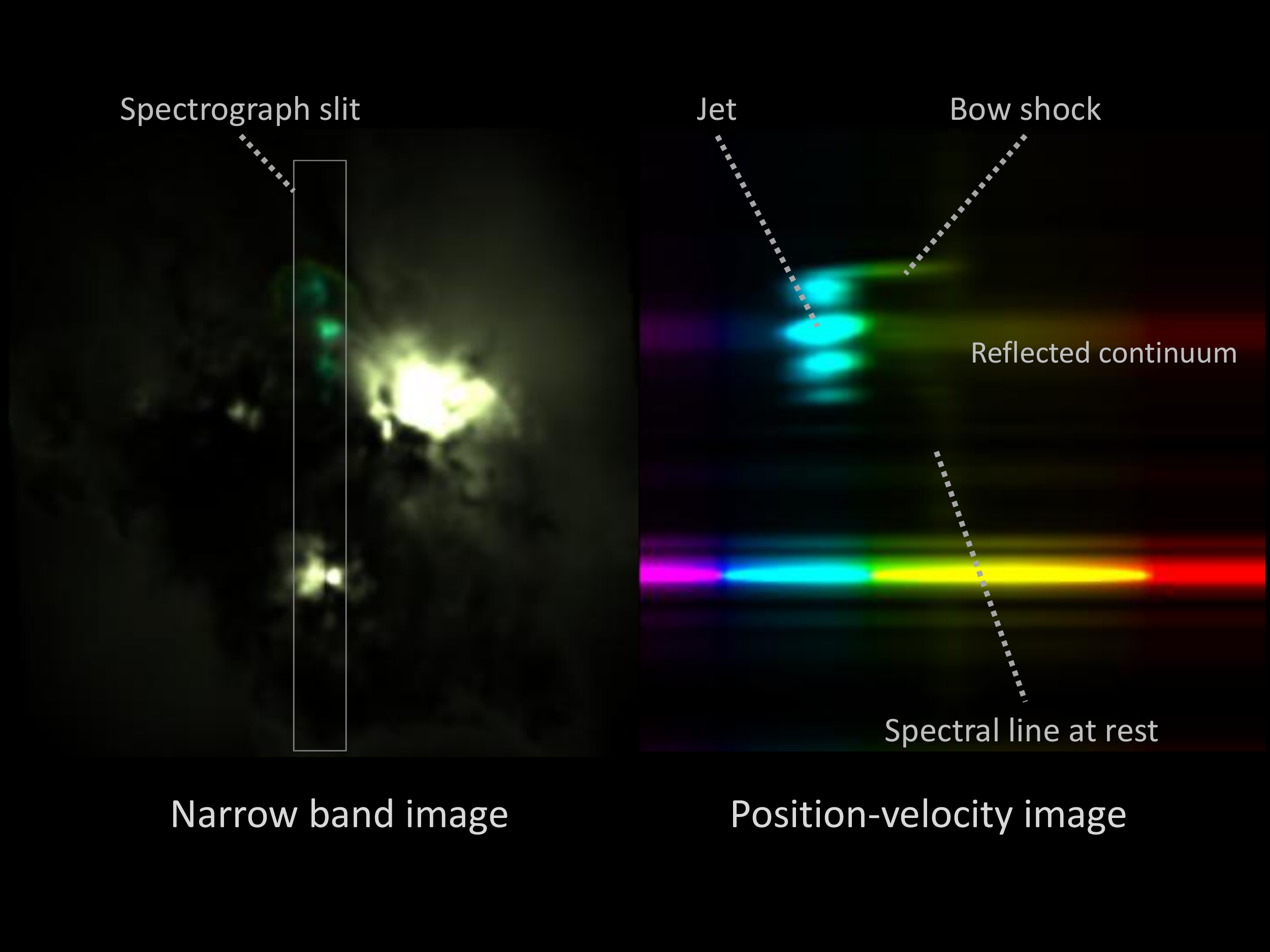}
\end{center}
\caption{The nebula model from Figure (\ref{polygonhydromix_nebula.fig}) is shown as seen through a narrow band filter around the emission line of the hydrodynamic jet. This allows the Doppler effect from the moving gas in the jet to be measured in the spectral line as a function of position along the synthetic spectrograph slit (along the jet). The picture on the right shows the position-velocity (P-V) image in a wavelength range that corresponds to a velocity range of $\pm 200$ km/s around the rest wavelength. The vertical direction is the position along the slit while the horizontal is the velocity component along the line of sight. This P-V image not only shows the influence of the gas motion, but also of the obscuring dust on the spectral line as the counter jet is mostly obstructed by dust. }
\label{longslit_nebula.fig}
\end{figure*}

In Figure (\ref{longslit_nebula.fig}) we show the result of changing only the spectral range of the rendering of the nebula from Figure (\ref{polygonhydromix_nebula.fig})with its hydro jet to a narrow range around the $H\alpha$ line emitted by the jet. On the left is an RGB broad-band image of the nebula with the jet. In the middle and on the right is the corresponding narrow-band image and the position-velocity (P-V) diagram. The absorption by the dust renders the P-V diagram asymmetric making a detailed interpretation problematic. A similar phenomenon would, of course, occur when obtaining P-V diagrams from the jet when partially embedded in the dusty star forming region. The P-V diagrams were derived using a synthetic spectrograph slit that was positioned along the projected mean axis of the jet, i.e. centered and vertical in the image in the middle.

Note that the simple change of the spectral range allows the switch from a visualization with educational content for the general public to a scientific model, that reveals the kinematics of the jet as seen through the absorbing regions of the dusty nebula. The absorption becomes evident through the asymmetry in the brightness of the jet compared to the counter-jet, both in the images and in the position-velocity diagrams.

Such complex models of a star forming nebula with multiple stars and jets can be used to study the observational challenges of measuring the kinematics of jets that are
embedded in complex dusty nebulae. Clearly, interpreting the detected position-velocity diagram of the jet section that is unobstructed by the dust is easier than the occluded counter jet, since all sections of the jet are visible when they are in front of the dust. But in addition to the kinematics of the jet, the continuum spectrum of the different regions can be studied as a function of the position and absorbing dust properties.

\section{Discussion and conclusions}
\label{discussion.sec}

In this paper we have shown how photorealistic astrophysical models can be obtained using interactive volumetric modeling based on polygon objects, hydrodynamic simulations and physical radiative transfer. The key new feature for the area of astrophysical simulation is the combination of polygon models and hydrodynamic simulations in a single model scene to enhance realism and scientific accuracy compared to pure polygon or pure hydrodynamic models. The result is beneficial not only for scientific analysis
and modeling of observations, but also for educational purposes be it in the context of outreach to the general public, school or university teaching.

The techniques that we have developed as part of the interactive virtual astrophysical laboratory \emph{Shape} allow not only the rendering still or animated space scenes in the visual spectral range, but also for an extended range of wavelengths with a single model. This is done through the physical description of the gas state as a function of position and wavelength. Improved visual and scientific realism is achieved through incorporation of hydrodynamic simulations using the corresponding software module or by importing external simulation data.

One of the educational values of such models to the general public is that they make it easier to convey the importance of long-wavelength observations in present terrestrial observatories and especially in the era of the \emph{James Webb Space Telescope}. But also to professional astronomers and their students, visualizations of this type, especially when the view point is animated in a movie or in an interactive visualization system, can be helpful in the process of exploration of the effects that might cause unexpected observational results. Naturally, the models can be applied as educational illustrations during university lectures and scientific presentations.

Since the models can be exported as volumetric data cubes in a format that is suitable for rendering on General Purpose Graphics Processing Units (GPGPU), they can and have been used for interactive visualization, for example in planetarium systems. Several models have been produced for incorporation in such systems.

In conclusion, the combination of 3D polygon models with hydrodynamic models and physical radiation transfer can be applied to obtain structurally very detailed models that are suitable for scientific analysis in astrophysics as well as photorealistic models for educational purposes.

\section*{Acknowledgments}
W.~S. and N.~K. acknowledge support through grant UNAM-PAPIIT IN101014 and IN104017.

\end{document}